\documentclass[titlepage,12pt]{article}
\usepackage{amssymb,psfig,epsfig}
\def\shat{\hat{s}}

\def\sp{{\mathbf s}_1}
\newcommand{\sm}{{\mathbf s_2}}
\newcommand{\kh}{{\hat{\mathbf k}}}
\newcommand{\ph}{\hat{\mathbf p}}
\newcommand{\dhh}{\hat{\mathbf d}}
\renewcommand{\thefootnote}{\fnsymbol{footnote}}
\begin{document}
\begin{titlepage}
\vspace{2cm}
\begin{center}
{\LARGE {\bf Next-to-leading order QCD corrections to top quark spin
correlations
at hadron colliders: the reactions $q {\bar q} \to t{\bar t}  (g)$}}  \\
\vspace{2cm}
{\bf W. Bernreuther$^{a,}$\footnote{supported by BMBF contract 05 HT9 PAA 1},
A. Brandenburg$^{b,}$\footnote{supported by a Heisenberg fellowship of D.F.G.}
and Z. G. Si$^{a,}$\footnote{supported by a A. v. Humboldt fellowship}}
\par\vspace{1cm}
$^a$Institut f.\ Theoretische Physik, RWTH Aachen, 52056 Aachen, Germany\\
$^b$DESY-Theorie, 22603 Hamburg, Germany
\par\vspace{3cm}
{\bf Abstract:}\\
\parbox[t]{\textwidth}
{Future hadron collider experiments are expected to
record large to huge samples of $t\bar t$ events.
The analysis of these data with respect to $t\bar t$ spin-spin
correlations requires precise predictions of the production of these quark
pairs in a
general spin configuration. Therefore we have
computed, at next-to-leading order (NLO) in the QCD coupling, the spin
density matrices describing $t\bar t$ production 
by quark antiquark annihilation, which is
the dominant production
process at the Tevatron.  Moreover we have computed the strength of the
$t\bar t$ spin
correlation at NLO, using various spin quantization axes.
}
\end{center}
\vspace{2cm}
PACS number(s): 12.38.Bx, 13.88.+e, 14.65.Ha
\end{titlepage}
%
\setcounter{footnote}{0}
\renewcommand{\thefootnote}{\arabic{footnote}}
\setcounter{page}{1}
\noindent
Pairs of top and antitop quarks will be produced  copiously with the
upgraded Fermilab Tevatron
collider and in even larger numbers with the Large Hadron Collider (LHC) at
CERN, allowing for
detailed investigations of the properties
of these quarks in the future.  For this aim  the
study of top spin phenomena will play an important role.
This is because the top quark, as compared with lighter quarks,  is unique
in that it
is sufficiently
short-lived to prevent hadronization effects from diluting the
spin-polarisations
and spin-spin correlations that  were imprinted upon the $t$ and ${\bar t}$
quarks by their production mechanism. Hence
spin-polarization and spin-correlation phenomena will provide valuable
information about the
interactions of top quarks.
An attempt to detect these spin correlations in a small $t\bar t$ dilepton
sample
collected at the Tevatron was recently reported by the D0 collaboration
\cite{D0:2000}.
\par
Needless to say, in order to interpret future data, the predictions of
these spin phenomena
should be as precise as possible. Within the Standard Model (SM) top
antitop pair
production at hadron colliders is determined
by  QCD. So far the next-to-leading order (NLO) QCD corrections are known
 for the spin-averaged differential $t\bar t$ cross section
\cite{Nason:1988,Nason:1989,Beenakker:1989,Beenakker:1991}, while spin
correlations
 were analysed only to leading-order
\cite{Barger:1989,Brandenburg:1996,Stelzer:1996,Mahlon:1996,
Mahlon:1997,Chang:1996}
in ${\alpha}_s$.
In this letter we report on progress in providing a complete description of
these spin effects within QCD at NLO:
We have computed, as a first step in this direction, the $t\bar t$ spin
density matrices
for the parton reactions $q {\bar q} \to t{\bar t}, t{\bar t} g$ to order
$\alpha_s^3$.
We have also calculated, for these reactions, the degree of the ${t \bar t}$ spin correlation
at NLO for different $t$ and $\bar t$ spin quantization axes.
\par
At the Tevatron
$q\bar q$ annihilation is the dominant process for producing top quark
pairs while at the LHC
$t\bar t$ production is mainly due to gluon gluon fusion. The $t\bar t$
spin correlations
and spin polarisations can be inferred from appropriate angular correlations
and distributions
of the $t$ and $\bar t$  decay products. In the SM the main top decay modes are
$t\to b W \to b q {\bar q}',  b \ell \nu_{\ell}$. Among these final states the
charged leptons, or the jets from quarks of weak isospin -1/2 originating
from $W$ decay, are most sensitive to
the polarisation of the top quarks. Hence to leading order in the QCD
coupling one has to
treat the two parton reactions
\begin{equation}
gg, q{\bar q}  \rightarrow t{\bar t}   \rightarrow b {\bar b}  + 4f,
\label{eq:ttrec}
\end{equation}
where $f=q,\ell,\nu_{\ell}$. The amplitudes for these two- to six-body
processes, with intermediate
top quarks of non-zero total width, were
given first in \cite{Kleiss:1988}. The computation of the NLO QCD
corrections to these
amplitudes is quite a demanding task \footnote{At NLO, apart from the gluon
radiation reactions,  the processes
$g + q ({\bar q})\to t {\bar t} + q ({\bar q}) \to  b {\bar b}  + 4f + q
({\bar q})$
must also be taken into account.}. In view of the fact that the total width
$\Gamma_t$
of the top quark is much smaller than its mass, $\Gamma_t/m_t ={\cal
O}(1\%)$, one may however analyse these
reactions using the so-called narrow width or leading pole approximation
\cite{Stuart:1991,Aeppli:1994}. In our case this approach consists, for a
given process,
 in an expansion of the amplitude around the poles of the unstable $t$ and
$\bar t$ quarks, which amounts to
an expansion in powers of $\Gamma_t/m_t$. Only the leading term of this
expansion, i.e.,
the residue of the double poles is  considered here. Then the radiative
corrections
to (\ref{eq:ttrec}) can be classified into so-called factorisable and
non-factorisable
corrections. The non-factorisable NLO QCD corrections were calculated in
\cite{Beenakker:1999}. For the factorisable corrections
the square of the complete
 matrix element ${\cal M}^{(\lambda)}$ is
of the form
\begin{equation}
\mid {\cal M}^{(\lambda)}{\mid}^2 \propto {\rm Tr}\;[\rho
R^{(\lambda)}{\bar{\rho}}]
 = \rho_{\alpha'\alpha}
R^{(\lambda)}_{\alpha\alpha',\beta\beta'}{\bar{\rho}}_{\beta'\beta} .
\label{eq:trace}
\end{equation}
Here $R^{(\lambda)}$ denotes the density matrix for the production of on-shell
$t\bar t$ pairs, the label $\lambda$ indicates the process, and
$\rho,{\bar{\rho}}$ are the density matrices describing the decay
of polarised $t$ and $\bar t$ quarks, respectively, into specific final states.
The subscripts in  (\ref{eq:trace}) denote the  $t$, $\bar t$ spin indices.
Note that both the production and decay density matrices are gauge invariant.
\par
The one-loop QCD corrections to the semileptonic decays of polarised top
quarks and to
$t \to W + b$ can be extracted from the results of \cite{Czarnecki:1991} and
\cite{Schmidt:1996,Fischer:1999}, respectively. In the following we 
describe our
computation of the  density matrices for $t\bar t$ production by $q\bar q$
annihilation.
At NLO we have to consider the reactions
\begin{equation}
q(p_1) + {\bar q}(p_2) \rightarrow t(k_1) + {\bar t}(k_2),
\label{eq:qq}
\end{equation}
and
\begin{equation}
q(p_1) + {\bar q}(p_2) \rightarrow t(k_1) + {\bar t}(k_2) + g(k_3).
\label{eq:qqgluon}
\end{equation}
We define the production density matrix for the
 process (\ref{eq:qq})
 in terms of its transition matrix element as follows: (for ease of
notation we omit the
label $\lambda$ in the following)
\begin{equation}
R_{\alpha\alpha' ,\beta\beta'}=
\frac{1}{N_{qq}}
\sum_{{{\rm\scriptscriptstyle colors} \atop
{\rm\scriptscriptstyle initial}\;
{\rm\scriptscriptstyle spins} }}
\langle t_\alpha\bar t_\beta |{\cal T}|
\,q\bar q\,\rangle\;
\langle \,q \bar q\,|{\cal T}^\dagger|
t_{\alpha'}\bar t_{\beta'}\rangle\;,
\label{eq:Rdef}
\end{equation}
where
the factor $N_{qq} = (2N_C)^2=36$ averages over the spins and colors
of the initial $q\bar q$ pair.
The matrix structure of $R$  is
\par
\hfill\parbox{13.4cm}{
\begin{eqnarray*}
R_{\alpha\alpha',\beta\beta'}&=&
A \delta_{\alpha\alpha'}\delta_{\beta\beta'}
+B_{i} (\sigma^i)_{\alpha\alpha'}
\delta_{\beta\beta'} +{\bar B}_{i} \delta_{\alpha\alpha'}
(\sigma^i)_{\beta\beta'} \\
&&+\, C_{ij}(\sigma^i)_{\alpha\alpha'}
(\sigma^j)_{\beta\beta'}  \,\, ,
\label{eq:Rstruct}
\end{eqnarray*} }\hfill\parbox{0.8cm}{\begin{eqnarray}  \end{eqnarray} }
\par\noindent
where $\sigma^i$ are the Pauli matrices. Using rotational invariance the
`structure functions'  $B_i,{\bar B}_i$ and
$C_{ij}$ can be
further decomposed.   The function
$A$, which determines the ${\bar t}t$ cross section, is known
to next-to-leading order in $\alpha_s$ from the work of
\cite{Nason:1988,Beenakker:1991}. Because
of parity (P) invariance  the vectors  ${\bf B},{\bf\bar B}$ can have,
within QCD,  only a component normal to the scattering plane. This component,
which amounts to a normal polarisation of the $t$  and $\bar t$ quarks,
 is induced by the absorptive part of the scattering amplitude,
and it was computed for $q\bar q$ and $g g$ initial states in
\cite{Bernreuther:1996,Dharmaratna:1996}
to order $\alpha^3_s$.  The normal polarisation is quite small, both for
$t\bar t$ production
at the Tevatron and at the LHC. Parity and CP invariance of QCD dictates that
the functions $C_{ij}$, which encode the  correlation  between the  $t$ and
${\bar t}$ spins,
have the structure \cite{Bernreuther:1994}
\begin{equation}
 C_{ij} = c_1\delta_{ij} + c_2
{\hat p}_{1i}{\hat p}_{1j} + c_3
{\hat k}_{1i}{\hat k}_{1j} + c_4
({\hat k}_{1i}{\hat p}_{1j} + {\hat p}_{1i}{\hat k}_{1j}) ,
\label{eq:cij}
\end{equation}
where ${\ph}_1$ and ${\kh}_1$ are the directions of flight of the
initial quark  and of the $t$ quark, respectively, in
the parton c.m. frame. The production density matrix for the reaction
(\ref{eq:qqgluon})
can be defined and decomposed in an analogous fashion.

To Born approximation the  functions $c_r$ were given, e.g., in
\cite{Brandenburg:1996}.
In order to determine these functions to
order $\alpha_s^3$  we first computed the one-loop diagrams that contribute to
(\ref{eq:Rdef}). Dimensional regularization was employed to treat both the
ultraviolet
and the infrared and collinear singularities which appear in the diagrams.
The ultraviolet singularities were removed by
using the $\overline{\rm{MS}}$ prescription for the QCD coupling $\alpha_s$
and the on-shell definition of the top mass $m_t$. The initial quarks are
taken to be
massless. After renormalisation the density matrix for the $t\bar t$ final
state still contains single and double poles in $\epsilon = (4-D)/2$ 
due to soft and collinear singularities.
These poles are cancelled after including the contributions of the reaction
(\ref{eq:qqgluon}) and  mass factorization. For the latter we
used the $\overline{\rm{MS}}$ factorization scheme. 
We avoided the computation of the exact density matrix for
the reaction (\ref{eq:qqgluon}) in $D$ dimensions by employing a
simple version of the phase-space slicing
method \cite{Giele:1992,Giele:1993}: We divided the phase space
into four regions, namely the region where the gluon is soft, the two 
regions where the gluon is  
collinear to one of the initial state massless
quarks (but not soft), and the complement of these three regions, where
all partons are `resolved'.
This decomposition can be performed using a single dimensionless 
cut parameter $x_{\rm min}$. For example, the soft region is defined
by the requirement that the scaled gluon energy
in the c.m. system $x_g=2E_g/\sqrt{s}$ is smaller than $x_{\rm min}$.
 In the soft region we used the eikonal
approximation of the matrix element for reaction (\ref{eq:qqgluon}) and 
the soft limit of the phase space measure. The integration over the
gluon momentum can then be carried out analytically in $D$ dimensions. 
The two collinear regions are defined by 
$(\cos\theta_{qg}>(1-x_{\rm min})$ and $ x_g>x_{\rm min})$
and  $(\cos\theta_{qg}<(-1+x_{\rm min})$ and $ x_g>x_{\rm min})$, 
respectively, where
$\theta_{qg}$ is the angle between the gluon and the quark in the 
$q\bar{q}$ c.m. frame. In these regions we used the collinear 
approximations for both the squared matrix element and 
the phase space in $D$ dimensions.
Finally, the exact spin density matrix  for reaction (\ref{eq:qqgluon})
in four space-time dimensions was used in the resolved region, where
all necessary phase space integrations can be carried out numerically.
By construction, all four individual contributions depend 
logarithmically on the slicing parameter $x_{\rm min}$, but in the
sum only a residual linear dependence on $x_{\rm min}$ remains, which is due
to the approximations made in the soft and collinear regions.
By varying $x_{\rm min}$ between
$10^{-3}$ and $10^{-8}$ we checked
that for $x_{\rm min}\le 10^{-4}$ this residual dependence
is smaller than our numerical error (which is less than a permill
for all results discussed below). 
\par
After mass factorisation we are left with finite density matrices for the
$t\bar t$
and the $t\bar t$ + hard gluon final states. 
As a check of our calculation we first compute the total cross section for
$q\bar q\to t{\bar t} + X$ at NLO. 
If one identifies the $\overline{\rm{MS}}$ renormalisation scale $\mu$
with the mass factorisation scale $\mu_F$  and neglects all quark
masses except for $m_t$, then one can express the cross section in
 terms of dimensionless scaling functions \cite{Nason:1988}:
\begin{equation}
\hat{\sigma}_{q\bar q}(\shat,m^2_t)=\frac{\alpha_s^2}{m^2_t}[
f^{(0)}_{q\bar q}(\eta) + 4\pi\alpha_s(f^{(1)}_{q\bar q}(\eta) +
{\tilde f}^{(1)}_{q\bar q}(\eta) \ln(\mu^2/m^2_t))],
\label{eq:xsection}
\end{equation}
where $\shat$ is the parton c.m. energy squared and $\eta = \shat/4m^2_t -1$.
We have compared our result for ${\sigma}_{q\bar q}$ as a function of
$\eta$
with those of \cite{Nason:1988,Beenakker:1991} and found perfect agreement.

\par
We now consider the following set of spin-correlation
observables:
\begin{equation}
{\cal O}_1=4\,\sp\cdot\sm,
\label{eq:sbasis}
\end{equation}
\begin{equation}
{\cal O}_2=4\,(\kh_1\cdot\sp)(\kh_2\cdot\sm),
\label{eq:hbasis}
\end{equation}
\begin{equation}
{\cal O}_3=4\,(\ph_1\cdot\sp)(\ph_1\cdot\sm),
\label{eq:pbasis}
\end{equation}
\begin{equation}
{\cal O}_4=4\,(\ph_2^*\cdot\sp)(\ph_1^{**}\cdot\sm),
\label{eq:ybasis}
\end{equation}
\begin{equation}
{\cal O}_5=4\,(\dhh_1\cdot\sp)(\dhh_2\cdot\sm),
\label{eq:obasis}
\end{equation}
where $\sp,\sm$ are the $t$ and $\bar t$ spin operators, respectively. 
The factor of 4 is conventional. With this normalization, 
the expectation value of ${\cal O}_1$ is equal to 1 at the Born level.
The expectation values of the  observables
(\ref{eq:hbasis}), (\ref{eq:pbasis}), (\ref{eq:ybasis}), 
and (\ref{eq:obasis}) determine the
correlation of different $t,\bar t$ spin projections.
Eq. (\ref{eq:hbasis}) corresponds to a 
correlation of the $t$ and $\bar t$ spins
in the helicity basis, while (\ref{eq:pbasis}) correlates the spins projected
along the beam line in the parton c.m.s. 
The `beam-line basis' used in (\ref{eq:ybasis}) was defined
 in \cite{Mahlon:1996} and refers to spin axes being parallel to the
antiquark direction
in the $t$ rest frame $\ph_2^*$ 
and to the quark direction in the
$\bar t$ rest frame  $\ph_1^{**}$, respectively. The spin axes
$\dhh_{1,2}$ in (\ref{eq:hbasis}) correspond to the so-called `optimal
basis' \cite{Parke:1996,Mahlon:1997} to be discussed below.
\par
For quark-antiquark
annihilation
it turns out that the spin
correlation (\ref{eq:pbasis}) \cite{Brandenburg:1996,Chang:1996}
and the correlation in the beam-line basis (\ref{eq:ybasis})
\cite{Mahlon:1996} are  stronger  than the correlation in the helicity basis.
A spin-quantization axis was constructed in \cite{Parke:1996,Mahlon:1997}
with respect to which the $t$ and $\bar t$
spins are 100$\%$ correlated to leading order in the
QCD coupling,  for all energies and scattering angles.
In terms of the structure functions of (\ref{eq:Rstruct}) this means
that the `optimal' spin axis $\dhh$ fulfills the condition
\begin{equation}
\hat{d}_i C_{ij} \hat{d}_j = A.
\label{eq:optcons}
\end{equation}
The existence of a solution of Eq. (\ref{eq:optcons}) is a special
property of the leading order spin density matrix for 
the reaction $q\bar{q}\to t\bar{t}$. One finds \cite{Parke:1996,Mahlon:1997}:
\begin{equation}
\dhh = \frac{-\ph_1+(1-\gamma_1)(\ph_1\cdot\kh_1)\kh_1}
{\sqrt{1-(\ph_1\cdot\kh_1)^2(1-\gamma_1^2)}}
, 
\end{equation}
where $\gamma_1=E_1/m_t$.
The construction of this axis 
explicitly uses the leading order result for the spin density matrix, and
different generalizations to higher orders are possible. We use in
(\ref{eq:obasis}) as spin axes:
\begin{eqnarray}
& &\dhh_1=\dhh, \nonumber \\
& & \dhh_2 = \frac{-\ph_1+(1-\gamma_2)(\ph_1\cdot\kh_2)\kh_2}{\sqrt{1-(\ph_1\cdot\kh_2)^2(1-\gamma_2^2)}}
, 
\end{eqnarray}
where  $\gamma_2=E_2/m_t$. For the 2 to 2 process $q\bar{q}\to t\bar{t}$,
$\dhh_2=\dhh_1=\dhh$. (A different generalization of $\dhh$ to higher orders
was used in \cite{Kodaira:1999}.)
\par
The expectation value of a spin-correlation observable ${\cal O}$
at parton level can be written at next-to-leading order in analogy to (\ref{eq:xsection})
as follows:
\begin{equation}
\langle {\cal O} \rangle_{q\bar q}  =
g^{(0)}_{q\bar q}(\eta) + 4\pi\alpha_s(g^{(1)}_{q\bar q}(\eta) +
{\tilde g}^{(1)}_{q\bar q}(\eta) \ln(\mu_F^2/m^2_t)).
\label{eq:expval}
\end{equation}
Note that these quantities  depend explicitly 
on the factorization scale $\mu_F$, but only implicitly (through $\alpha_s$)
on the renormalization scale $\mu$. This is because a factor $\alpha_s^2$
drops out in the expectation values, and hence the Born result is of order
$\alpha_s^0$. 
\par
Our results for the functions 
$g^{(0)}_{q\bar q}(\eta),\ g^{(1)}_{q\bar q}(\eta)$ and
${\tilde g}^{(1)}_{q\bar q}(\eta)$ are shown for the five observables
(\ref{eq:sbasis})-(\ref{eq:obasis}) in Figs. 1-5. In each figure,
the dotted line is the Born result $g^{(0)}_{q\bar q}(\eta)$, the full
line shows the function $g^{(1)}_{q\bar q}(\eta)$, and the dashed
line is ${\tilde g}^{(1)}_{q\bar q}(\eta)$. A general feature of all 
results is that the QCD corrections are very small for values of
$\eta\lesssim 1$. For larger values of $\eta$, the functions 
$g^{(1)}_{q\bar q}(\eta)$ depart significantly from zero. Also, 
the functions ${\tilde g}^{(1)}_{q\bar q}(\eta)$ become nonzero, with
less dramatic growth as $\eta\to \infty$ and with an opposite sign
as compared to $g^{(1)}_{q\bar q}(\eta)$. The phenomenological 
implications of these features for spin correlations
at the Tevatron and the LHC will be studied in detail
in a future work. Here we merely note that the substantial QCD corrections
for large $\eta$ will be damped by the parton distribution functions,
which decrease rapidly with $\eta$. Moreover, at Tevatron energies values
of $\eta$ above $\sim 30$ are kinematically excluded.
\par
To summarize: We have computed the 
spin density matrices describing $t\bar{t}$ production 
by $q\bar{q}$ annihilation to order $\alpha_s^3$. Further we have evaluated
the scaling functions encoding the QCD corrections to 
spin correlations, using a number of different spin quantization axes.
This work provides a building block, which was missing so far,
 towards a complete description
of spin effects in the hadronic production of top quark pairs at NLO in
the strong coupling.

\subsubsection*{Acknowledgments}
We would like to thank M. Spira and P. Uwer for discussions.
Z. G. Si wishes to thank the DESY theory group for its hospitality
during the final stages of this work.

\newpage
\begin{center}
FIGURE CAPTIONS
\end{center}
\noindent 
{\bf Fig. 1.} Dimensionless scaling functions $g^{(0)}_{q\bar q}(\eta)$ 
(dotted), $g^{(1)}_{q\bar q}(\eta)$ (full), and 
${\tilde g}^{(1)}_{q\bar q}(\eta)$ (dashed) that determine 
the expectation value $\langle {\cal O}_1 \rangle_{q\bar q}$.\\
{\bf Fig. 2.} Same as Fig.1, but for $\langle {\cal O}_2 \rangle_{q\bar q}$. \\
{\bf Fig. 3.} Same as Fig.1, but for $\langle {\cal O}_3 \rangle_{q\bar q}$. \\
{\bf Fig. 4.} Same as Fig.1, but for $\langle {\cal O}_4 \rangle_{q\bar q}$. \\
{\bf Fig. 5.} Same as Fig.1, but for $\langle {\cal O}_5 \rangle_{q\bar q}$. 
\newpage
\begin{figure}
\unitlength1.0cm
\begin{center}
\begin{picture}(16,16)
\put(0,0){\psfig{figure=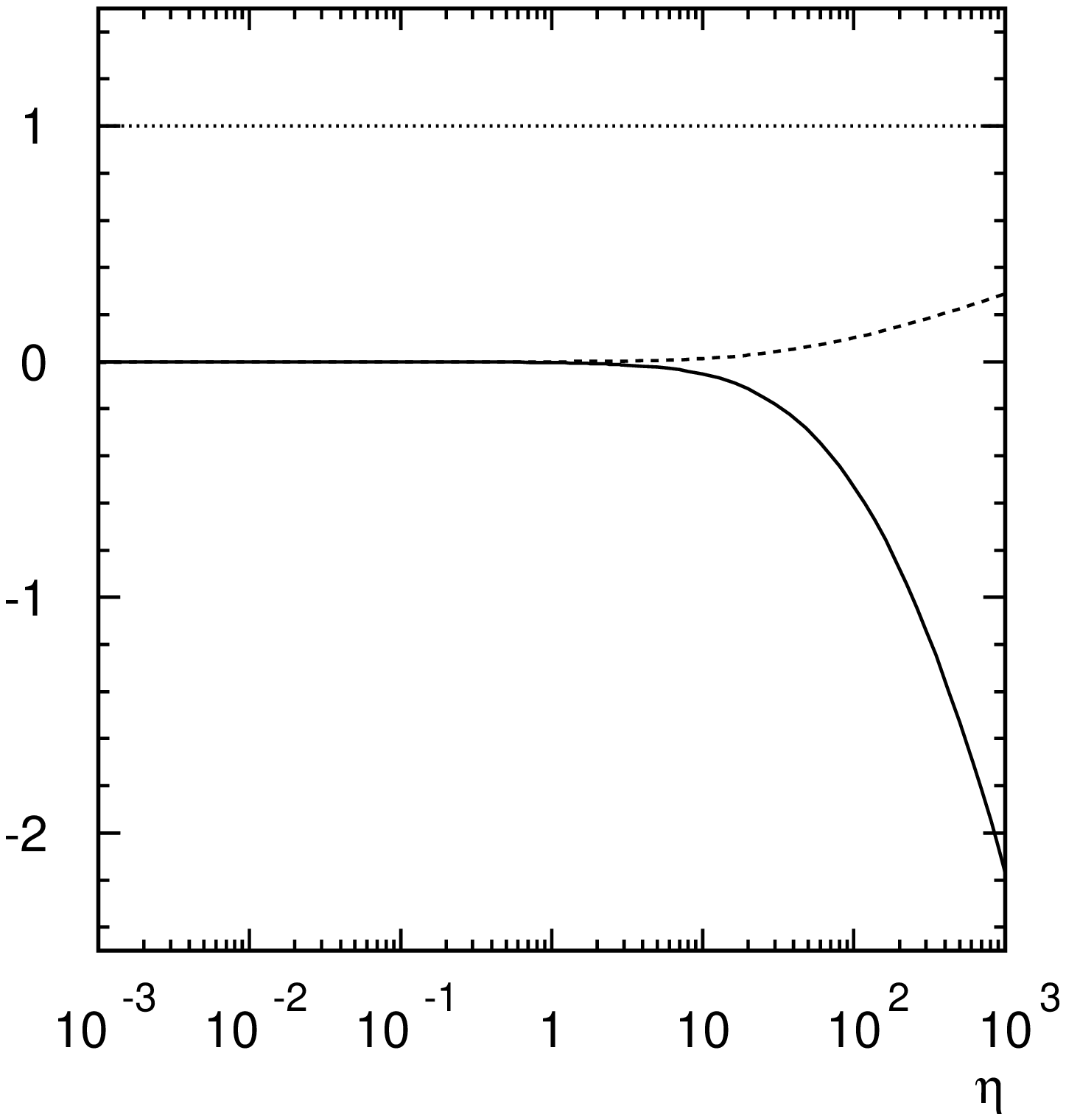,width=12cm,height=12cm}}
\end{picture}
\vskip 0.5cm
\caption{}\label{fig:obs1}
\end{center}
\end{figure}
\begin{figure}
\unitlength1.0cm
\begin{center}
\begin{picture}(16,16)
\put(0,0){\psfig{figure=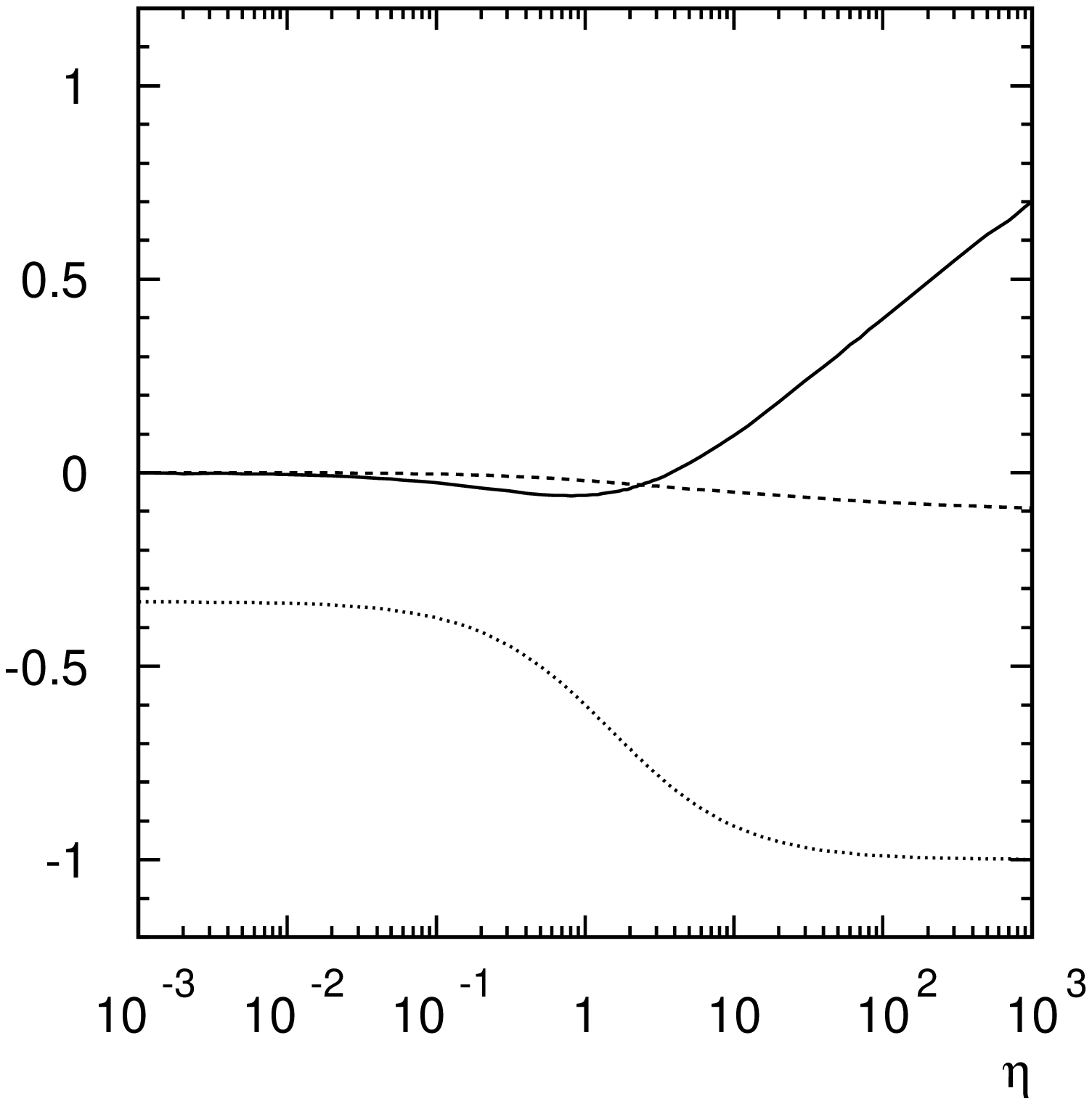,width=12cm,height=12cm}}
\end{picture}
\vskip 0.5cm
\caption{}\label{fig:obs2}
\end{center}
\end{figure}
\begin{figure}
\unitlength1.0cm
\begin{center}
\begin{picture}(16,16)
\put(0,0){\psfig{figure=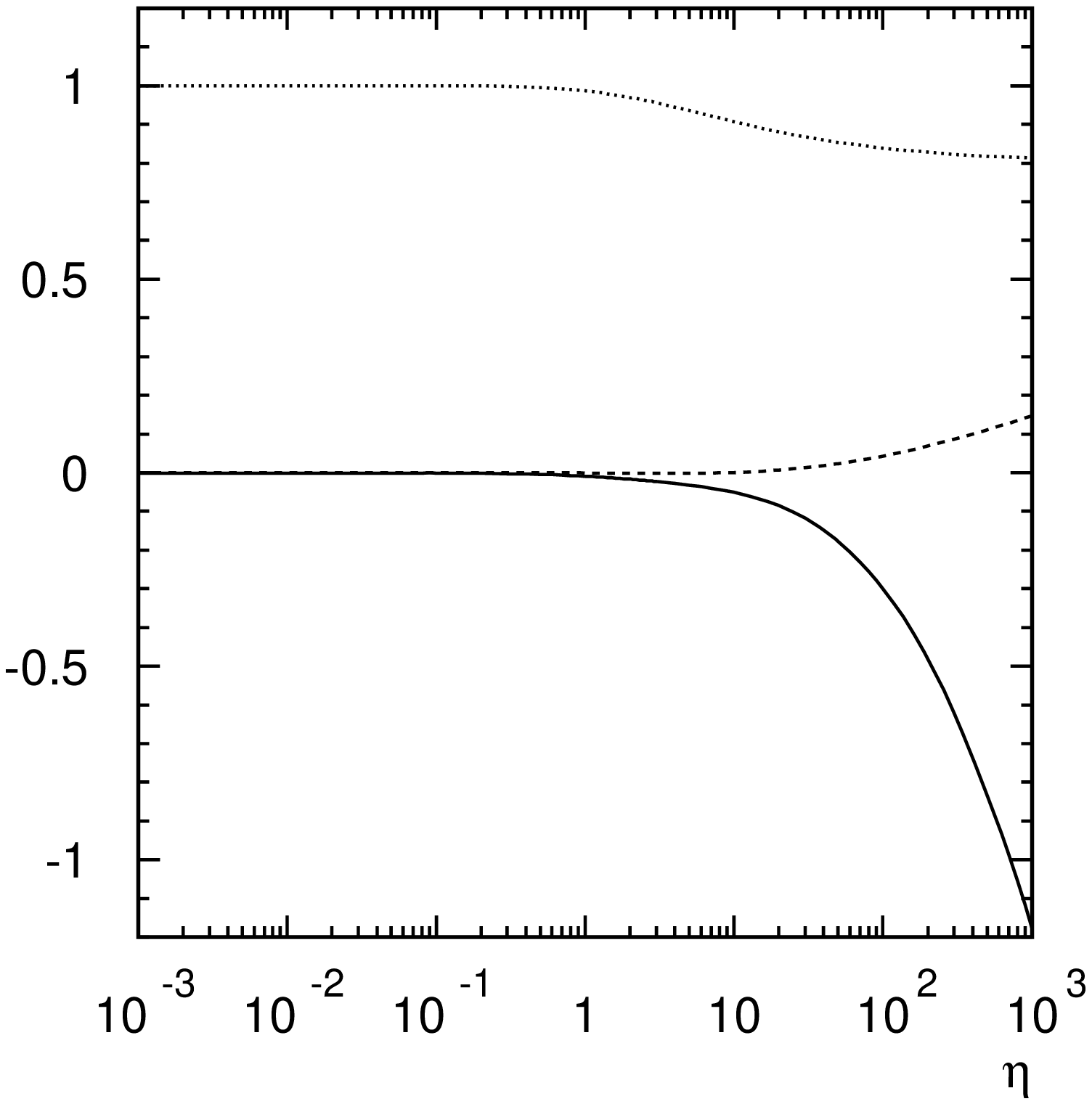,width=12cm,height=12cm}}
\end{picture}
\vskip 0.5cm
\caption{}\label{fig:obs3}
\end{center}
\end{figure}
\begin{figure}
\unitlength1.0cm
\begin{center}
\begin{picture}(16,16)
\put(0,0){\psfig{figure=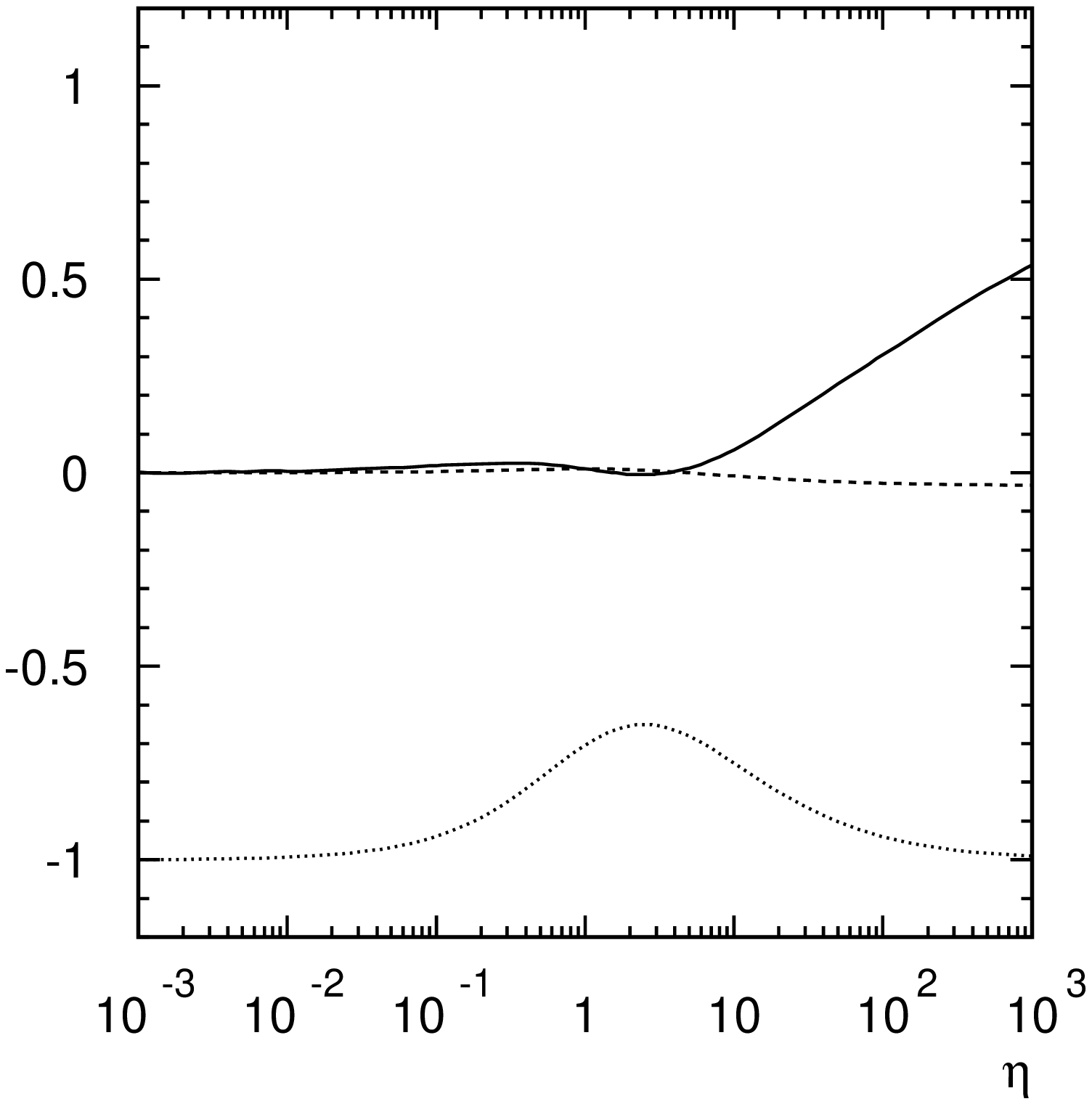,width=12cm,height=12cm}}
\end{picture}
\vskip 0.5cm
\caption{}\label{fig:obs4}
\end{center}
\end{figure}
\begin{figure}
\unitlength1.0cm
\begin{center}
\begin{picture}(16,16)
\put(0,0){\psfig{figure=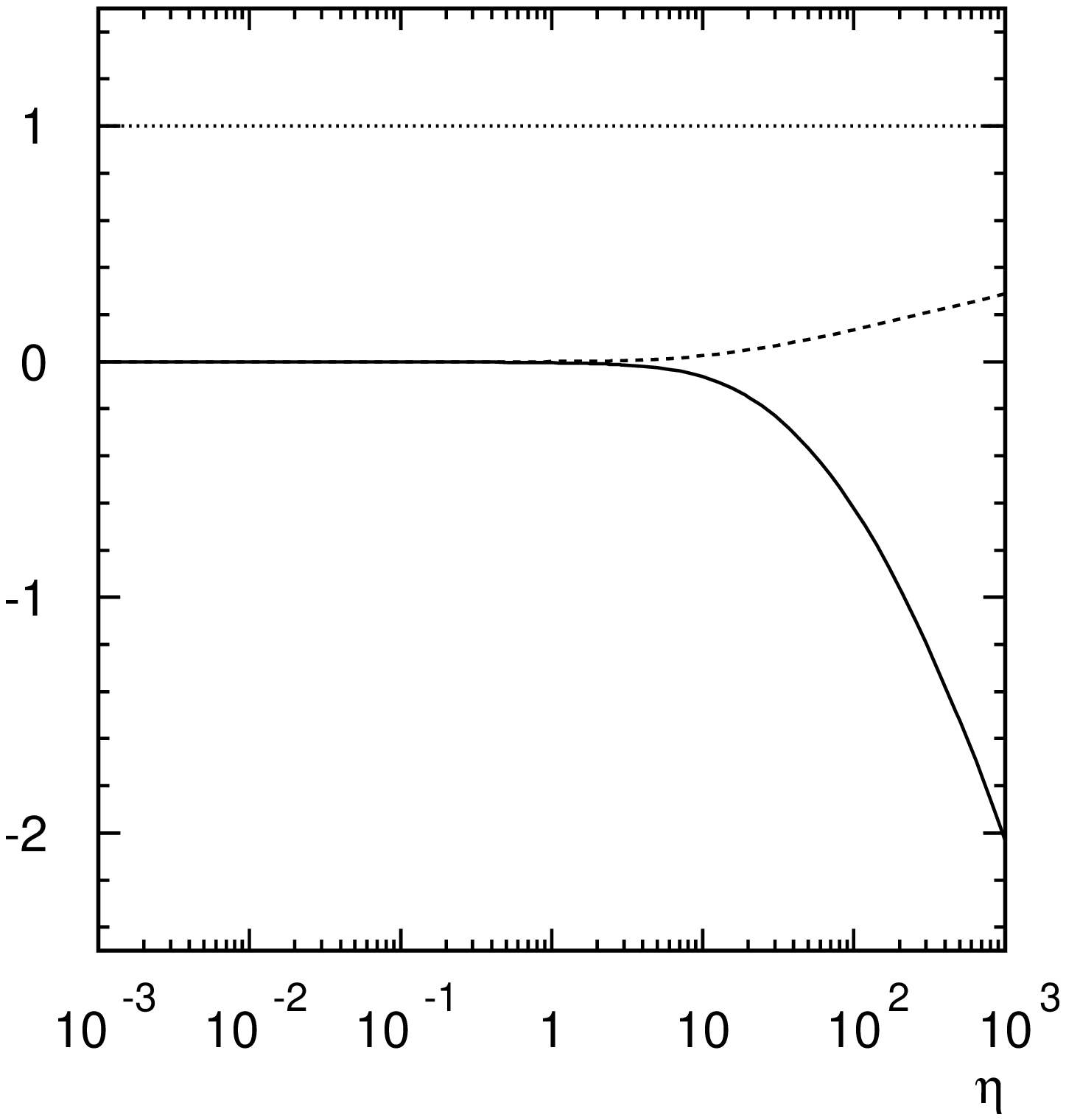,width=12cm,height=12cm}}
\end{picture}
\vskip 0.5cm
\caption{}\label{fig:obs5}
\end{center}
\end{figure}
\end{document}